\documentclass[a4paper]{elsarticle}
\usepackage{graphicx}
\usepackage{color}
\def\Vec#1{\mbox{\boldmath $#1$}}

\begin{document}
\begin{frontmatter}
\title{
Molecular Origin of Limiting Shear Stress of Elastohydrodynamic Lubrication Oil Film Studied by Molecular Dynamics}

\author[tytlabs,uhyogo,kyoto-u]{Hitoshi Washizu\corref{cor1}}
\ead{h@washizu.org}
\author[tytlabs]{Toshihide Ohmori}
\author[aisin]{Atsushi Suzuki}
\address[tytlabs]{Toyota Central R\&D Labs., Inc., Nagakute, Aichi 480-1192, Japan}
\address[uhyogo]{Graduate School of Simulation Studies, University of Hyogo
  7-1-28 Minatojima-minamimachi, Chuo-ku, Kobe, Hyogo, Japan}
\address[kyoto-u]{Elements Strategy Initiative for Catalysts and Batteries,
Kyoto University, Katsura, Kyoto 615-8520, Japan}
\address[aisin]{AISIN AI Corporation, 1-33 Kounosu, Nishiazai-Cho, Nishio, Aichi, Japan}
\cortext[cor1]{Corresponding author}

\begin{abstract}
All-atom molecular dynamics simulations of an elastohydrodynamic
lubrication oil film are performed to study 
the effect of pressure.
Fluid molecules of
n-hexane are confined between two solid plates under a constant normal
 force of 0.1--8.0 GPa.
Traction simulations are performed by applying relative sliding
motion to the solid plates. A transition in the traction
behavior is observed around 0.5--2.0 GPa, which corresponds to the
viscoelastic region to the plastic--elastic region, which are
experimentally observed. 
This phase transition is related to the
suppression of the fluctuation in molecular motion.
\end{abstract}

\section*{Highlights}

\begin{itemize}
 \item A mechanism for the 
      limited shear-stress transition of an elastohydrodynamic lubrication
      (EHL) oil film is proposed.
 \item The spatial structure and velocity profile showed gradual changes.
 \item Suppression of the fluctuation is found in the EHL regime.
\end{itemize}

\end{frontmatter}


\section{Introduction}

Machine elements in which large loads are transmitted, such as a traction-drive continuously variable transmission (CVT), work in the
elastohydrodynamic lubrication (EHL) regime~\cite{Dowson:1977}.
The molecular dynamic
behavior of the oil film for EHL is not well-understood since
the long trajectories of ensembles of a large number of fluid molecules are
required to analyze the drastic phase transition induced by a high pressure.
Typical practical EHL conditions include a film thickness on the order of a
submicron and a shear rate less than 10$^6$ /s for 
a submicron thick film~\cite{Washizu:2010}. 
Although this film
thickness exceeds the range of van der Waals structural forces of the
solid plates, the behavior of the fluid layer differs from that of the bulk,
i.e., the traction coefficients (the tangential traction force divided by
the normal load) cannot be easily deduced from the bulk shear
viscosity.

In this letter, we report molecular dynamics (MD) simulations
to study the effect of pressure.
The hydrocarbon molecules are confined between two solid atom layers,
and the pressure is added as a normal force to the solid atom layers.
The solid atom layers move at a constant velocity so that shear is
induced in the fluid film.
Previous simulation of a submicron-thickness oil film showed  
quantitative agreement between experiment and
simulation~\cite{Washizu:2010, Washizu:2014b}.
Here, we extend our study to examine the effect of an external pressure,
in molecular level.
In an EHL oil film,
for an increasing external pressure, 
traction, i.e., the friction coefficient, saturates
and does not depend on the shear rate
at some pressure, which is called
the limiting shear stress. 
Although this transition is assumed to have a relationship
with glass transition from
\textcolor{red}{liquid-like state to
solid-like state},
the state of the fluid in EHL film is not the
same as in a static condition of materials
{~\cite{Bair:2007}.
Moreover, 
the mechanism of the glass transition is under discussion,
especially for hydrocarbons. For example,
mode-coupling theory is a candidate for
explaining the glass transition of particles,
and the internal degrees of freedom coupled with the intermolecular
interaction are very complex to model~\cite{Miyazaki:2004}.
An all-atom
non-equilibrium MD simulation is an effective tool
to simulate
the phase behavior under shear
for understanding these phenomena. 
We use $n$-hexane fluid molecule as a typical
hydrocarbon oil, in order to obtain a universality
of the origin of limiting shear stress.

\section{Method}

All-atom
MD simulations of
a lubricating oil film confined between
solid walls under shear conditions
are executed using the following procedure.
$n$-hexane is chosen as the fluid molecule so that
the general behaviors of the hydrocarbon oils 
are simulated.
The related sliding speed of the solid walls
is set to $2 v_\mathrm{wall}$~=~1.0~m/s, which is the usual driving
condition in macroscopic machine elements
such as a CVT.

In the MD simulation,
the hydrocarbon molecules are dynamically treated using 
the AMBER force field~\cite{Kollman:1995}. 
On the $x, y$ 
and $z$ axes, periodic boundary conditions are adopted
for the hexahedron simulation cell, where
$x$ is the sliding direction, and $z$ is the direction of the
fluid film thickness.  
The temperature is controlled at 350~K using the
Nos\'e--Hoover thermostat~\cite{Nose:1984}. 
All molecules are connected to a heat bath. 
While this has been shown to affect the rheology of the confined fluids
at shear rates above $\dot\gamma = 0.2$
(in reduced Lennard--Jones units), 
the low shear
rates here, a maximum of 
$\dot\gamma (\epsilon / mR^{2})^{1/2} \approx 0.004$, 
suggest that such artifacts should be negligible
~\cite{Khare:1997a, Greenfield:2005}.
Each simulation step corresponds to 0.5~fs, and the time
integral due to the motion of the atoms and molecules 
is calculated by the 
reversible reference system propagation algorithm (rRESPA)
~\cite{Tuckerman:1992} 
method. The solid plate of an
alpha-ferrous crystal is modeled as a solid atom layer with a lattice
of $10 \times 10 \times 3$ 
in the $x, y, z$ directions, and the lattice parameter is set to 2.87 \AA. 
The fluid--solid interface is the (100) surface, and the
vibrations of each solid atom are suppressed.  
The Lennard--Jones parameter for solid atoms is then set to 
12.14 kcal/mol~\cite{Washizu:2010} 
in order to suppress boundary slip~\cite{Washizu:2014b}.

Constant shear states under a constant pressure and constant 
sliding velocity are obtained using the following procedure. 
A set of fluid
molecules is first arranged in a lattice configuration. 
The molecules are
then moved by the MD simulation under the periodic boundary conditions in
the $x, y$, and $z$
directions at a constant temperature of 350~K until a thermal
equilibrium is obtained. 
Next, several sets of thermal equilibrium fluids
are arranged in the $z$ direction between two solid plates. 
Fluid molecules are compressed by the MD simulation under the periodic
boundary conditions in the $x$ and $y$ directions by applying a constant pressure,
$p_{zz}$, to the plates. 
The stress in the plates is in the $z$ direction.  
A shear in
the $x$ direction is then applied by the relative sliding motion of the Fe
plates at the constant velocity, $v_\mathrm{wall}$.
The number of atoms is set to 576,
which corresponds to a film thickness of 10~nm at 1~GPa 
to simulate a sufficiently thick oil film.
If the oil film thickness is too small,
the solid like behavior, such as stick-slip occur~\cite{Washizu:2010}.
The film thickness is calculated from the average
of the difference between the highest and lowest fluid molecules in the fluid film.
The simulation code is parallelized
and tested on a massively parallel computer,
which was described in our
previous work~\cite{Washizu:2010, Washizu:2014b}.

\section{Results and Discussion}

The effect of the external pressure is studied by 
simulations of a 50-ns-long trajectory for each $p_{zz}$. 
Snapshot of the simulation of an $n$-hexane film 
in steady shear state with 
the relative sliding speed of $2 v_\mathrm{wall}$~=~1.0~m/s,
and at external pressure $p_{zz} = 8.0$~GPa, is  shown in Fig. 1.
The highly layered fluid molecules are found
in the vicinity of the wall which is already reported
in our previous work~\cite{Washizu:2010, Washizu:2014b}.
In comparison with snapshots in other external pressures
(they are shown in the graphical abstract),
the film thickness is small and the density of the
fluid molecules is high. 

Figure~2a shows the external pressure dependence of
the film thickness.
The film thickness decreases monotonically with the increase in
the external pressure.
This tendency resembles that observed in experiments
~\cite{Bridgman:1949}.
\textcolor{red}{
  In more detail, there would be a step decrease in the
  film thickness, which is shown for the liquid-solid transition
  in the bulk. In our simulation, however, finding the precise
  transition point is out of scope, since the system is under
  confinement and shear, we cannot compare with the experiment
  directly. The comparision of the static phase transition
  with the experiments in many kind of fluid molecules will
  be shown in the our next paper.}
Figure~2b shows the external pressure dependence of the traction coefficient.
The traction coefficient is calculated from the average force
acting on both solid sheets divided by the external pressure $p_{zz}$. 
The traction coefficient first increases monotonically with the
increase in the external pressure and plateaus
when $p_{zz} \ge 1.0$~GPa.
Although the shear rate differs in our simulation since
the number of molecules is fixed so that the film thickness
decreases owing to the increase in the external pressure,
the simulation results show saturation of the shear stress.
The limiting shear stress observed in the experiments is
reproduced in the simulation, and 
a transition in the change in the traction coefficients is
observed around  $p_{zz}$~=~0.5--1.0~GPa.
This pressure regime may correspond to 
the melting pressure of $n$-hexane at 350K~\cite{Morawski:2005}.
Although
the range of critical $p_{zz}$ almost corresponds to the
experimental range, the critical traction coefficient
cannot predicted quantitatively since the film thickness
is very different under the experimental conditions and
simulation conditions~\cite{Washizu:2010}.
Therefore, we discuss the origin of the limiting shear
stress qualitatively in the following.

From the experiments, this transition corresponds to a
phase transition from the viscoelastic region
to the plastic--elastic region~\cite{Bair:2007}.
In order to analyze the molecular mechanism of the phase transition,
radial distribution functions (RDFs) are suitable for describing the
static structural change in the ensemble of fluid molecules.
Figure~3 shows the RDFs 
of the carbon atoms in the fluid molecules 
at the center of the fluid film.
The molecules are chosen such that they are located 25\% to 75\%
of the distance between plates.
Note that the first and second peaks correspond to the
peaks between atoms in their own molecules.
The outer region (r $>$ 0.3 nm) shows
the peaks in the intermolecular distribution of carbon atoms.
In the RDFs, liquid-like distributions that have
a single peak or two peaks in the outer region
are found at a low pressure of $p_{zz}$ = 0.1--0.5 GPa. 
As the pressure increases, the sharpness of the peaks
in the outer region increases, and
shifts in the peaks are observed.
This is due to the increase in the density of the oil film.
In other words, a decrease in the free volume of hydrocarbon molecules
occurs, and the internal bond length between each internal atom
is preserved within the simulated range of external pressures.
From the viewpoint of the phase transition, however, 
the change in the RDFs is gradual, and 
an obvious transition in the
curves around $p_{zz}$ = 1.0 GPa is not observed. 
This means that the transition is not
only related to the arrangement of the molecules.
 
In order to discuss the dynamic motion of the molecules,
the velocity profiles at
a low pressure ($p_{zz}$ = 0.5 GPa), high pressure 
($p_{zz}$ =  2.0 GPa), and very high pressure
($p_{zz}$ =  8.0 GPa) are
plotted in Fig. 4a.
$v_{x}, v_{y}$, and $v_{z}$ are the velocities
in the $x, y$, and $z$ directions, respectively,
and $<>$ denotes the time average.
At every external pressure $p_{zz}$, 
$<v_{y}>$ and $<v_{z}>$ are very low owing to the
conservation of momentum. 
The small oscillation of $<v_{z}>$ 
observed in the vicinity of both walls
may due to the structuring of the fluid
molecules~\cite{Washizu:2010, Washizu:2014b}.
The velocity profiles for $<v_{x}>$ differ for the
external pressures. 
Newtonian-like profiles are observed at a low
pressure, whereas non-Newtonian profiles are observed more strongly
at higher pressures,
which suggests an increase in the elastic interactions between
molecules. 
The difference, however, is to small to explain the
dramatic change in the traction behavior.

The differences in the molecular
motion at low and high pressures
are clearly shown in the profiles of
the velocity fluctuation, which are shown in Fig.~4b. 
At a low pressure of $p_{zz}$=0.5 GPa, the degree of
fluctuation in the velocity is three times larger than 
that at a high pressure.
Moreover, 
plateau regions, i.e.
regions in which the fluctuation curves are saturated
are observed at the center of the oil film.
This means that the effect of the wall is in a limited region
from the wall, and the motions of the molecules
do not depend on the film thickness.
The $(v_{x} - <v_{x}>)^2$ and $(v_{y} - <v_{y}>)^2$ curves are
almost identical and exhibit larger values than the $(v_{z} - <v_{z}>)^2$ curves.
This means that the fluctuations in the molecules are
suppressed
stronger
in the $z$ direction
than in the $x, y$ directions
owing to the system confinement.
Thus, the molecules can change positions in either the $x$ or $y$ directions,
which is described by the Eyring viscosity~\cite{Eyring:1936}.

At high pressures of $p_{zz}$=2.0 and 8.0~GPa,
a plateau region is not observed, which indicates that
the effect of the wall,
reaches the center of the film in
EHL regime.
This is not because 
the local pressure at the center of the film change as the normal pressure increase.
The region of the correlation, such as the momentum correlation length
in our previous paper~\cite{Washizu:2014b}, 
which is more than 50~nm under $p_{zz}$=1.0~GPa,
has changed due to the increase of the external pressure.
Here, the curves of $(v_{y} - <v_{y}>)^2$ and $(v_{z} - <v_{z}>)^2$
are identical. This means that fluctuation in the $y$ direction
is suppressed, as in $z$ direction, so that the molecules
are caught at fixed positions.
In the $x$ direction, the shear field causes a larger
fluctuation than in the other directions. 
The bump shown in the curve of 
$(v_{x} - <v_{x}>)^2$ at $p_{zz}$~=8.0~GPa may be
related to the larger shear field at the center of the oil film.
This suggests that the
plasticity is related to the suppression of fluctuation in the
molecular motion.

This tendency is also observed from an analysis of the segment (bond) motions.
In Fig.~5, the autocorrelation functions 
$P_1(t) = <s(t) \cdot s(0)> = <\cos(\theta (t))>$
of each segment of the $n$-hexane molecules $\Vec s$ 
are plotted for the ensemble of molecules in the vicinity of the solid walls
and at the center of the fluid layers at
a low pressure of $p_{zz}$ = 0.5 GPa and high pressures of
$p_{zz}$ = 2.0 and 8.0 GPa.
Here, $s (t)$ is a unit vector of $\Vec s$, and
$\theta (t)$ is the angle between $\Vec s$ at times $t_0$ and $t_0 +t$.
Long-time tails of the correlation functions are observed in the vicinity of the walls.
The relaxation times of the segment motions calculated from fitting the curves 
at the center of the fluid are
312, 2,294, and 3,846 ps at $p_{zz}$ = 0.5, 2.0, and 8.0~GPa,
respectively.
At the wall, the relaxation times are
1,245, 10,913, and 23,161 ps at $p_{zz}$ = 0.5, 2.0, and 8.0~GPa,
respectively.
The longer relaxation times 
observed for higher pressures 
at the center of the films are consistent
with the explanation of the
velocity fluctuation in Fig.~4b. 
At high pressures of $p_{zz}$ = 2.0 and 8.0~GPa,
the shear field does not induce a relative
positional change in the molecular structure
in the $y$ and $z$ directions so that the correlations between
the bonds are almost fixed not only in the vicinity
of the walls but at the center of the film.

The MD simulations have demonstrated that
the EHL transition is not clearly observed from the change in the
static liquid structure to the solid-like structure shown in Fig.~3.
The dynamics of the average velocity in Fig.~4a
show slight differences.
A change is clearly observed for the velocity
fluctuation in Fig.~4b. The velocity fluctuation
is suppressed by a factor of more than three at a high pressure,
and the bulk region of the velocity fluctuation disappeared.
This change is also shown from the analysis of the segment motions
in Fig.~5.

We have not taken into consideration of occurence of shear bands
~\cite{Bair:2007}, which is one of the explanation to the
internal slip hypothesis for pressure independent shear stress. 
However, we have already obtained some characteristic features
of the limiting shear stress in EHL film.
From the continuous velosity profile in Fig. 4a, the shear localization
or slip, which is related to the shear band is not found.
In our simulation, the saturation of the traction
coefficient is related to the bump of the velocity fluctuation
at the center of the fluid film, which is shown 
in the $v_{x}$ curve of $p_{zz}$=~8.0 GPa in Fig.~4b.
The excess shear force may be canceled
by the increase of the fluctuation in $x$ direction at the center. 
The relation between our mechanism and experimentally observed
shear bands would be explained in the long time simulation in
realistic oil film thickness. 

In our simulation, the temperature is set to a constant
value of 350~K using the Nos\'e--Hoover thermostat for every
molecule in the system.
Although the shear softening effect by heating a viscoelastic
fluid by shear field~\cite{Kobayashi:2014} is not treated in
the simulation,
we already succeeded in reproducing the saturation in the
traction coefficient reported by experiments.

In order to explain all experimental phenomena,
the heat transport within the
film thickness under the experimental conditions
may also be studied in future simulations.

\section{Conclusions}

All-atom MD simulations are used to calculate the
effect of pressure under confinement and sliding. 
The limiting shear stress of the EHL
molecular oil film is reproduced by
the simulation. This phase transition is related to the suppression of
the fluctuation in the molecular motion.

\section*{Acknowledgement}

This work was partially
supported by the Next Generation Super Computing Project,
Nanoscience Program, Ministry of Education, Culture, Sports, Science and
Technology, Japan.  
We also thank Prof. Dr. Shi-aki Hyodo, Dr. Shuzo Sanda 
and Prof. Dr. Yuichi Nakamura for useful discussions.

\section*{References}


\newpage

\begin{figure}[h]
\begin{center}
 \includegraphics[width=40mm]{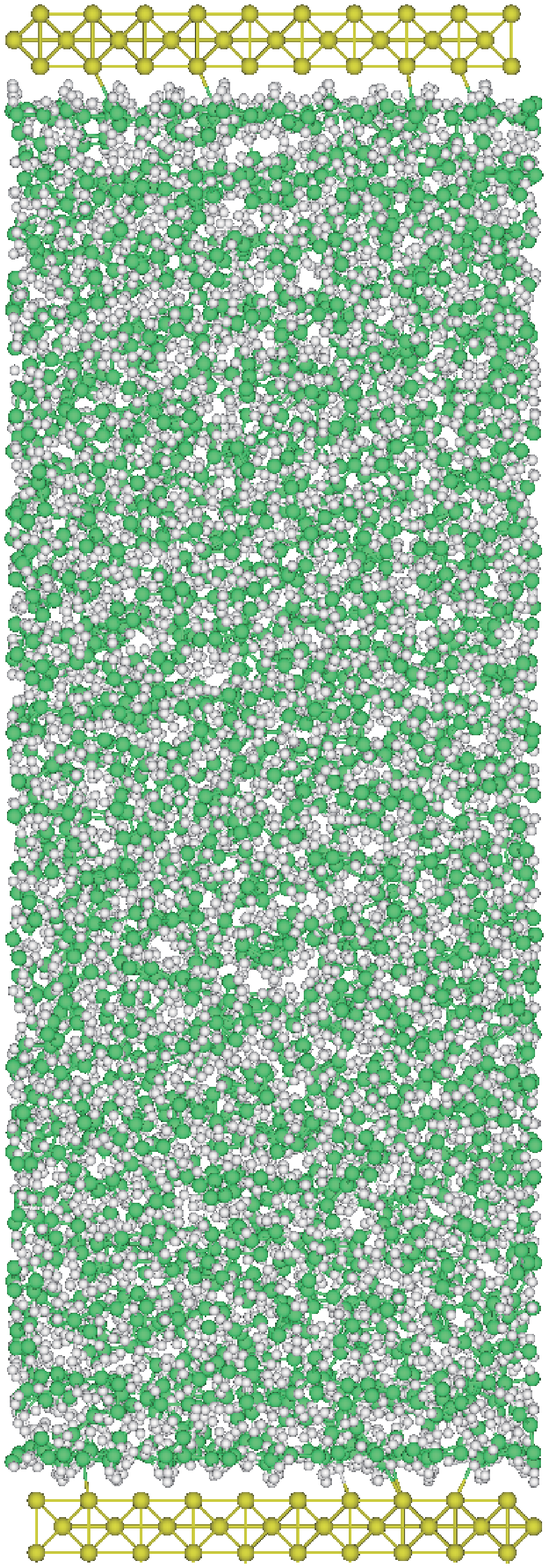}
 \caption{
 {
   Snapshot of the molecular dynamics simulation
   of n-hexane fluid film at
   external pressures $p_{zz} = 8.0$ GPa.
 }}
  \label{fig:p80}
\end{center}
\end{figure}

\begin{figure}[h]
\begin{center}
 \includegraphics[width=80mm]{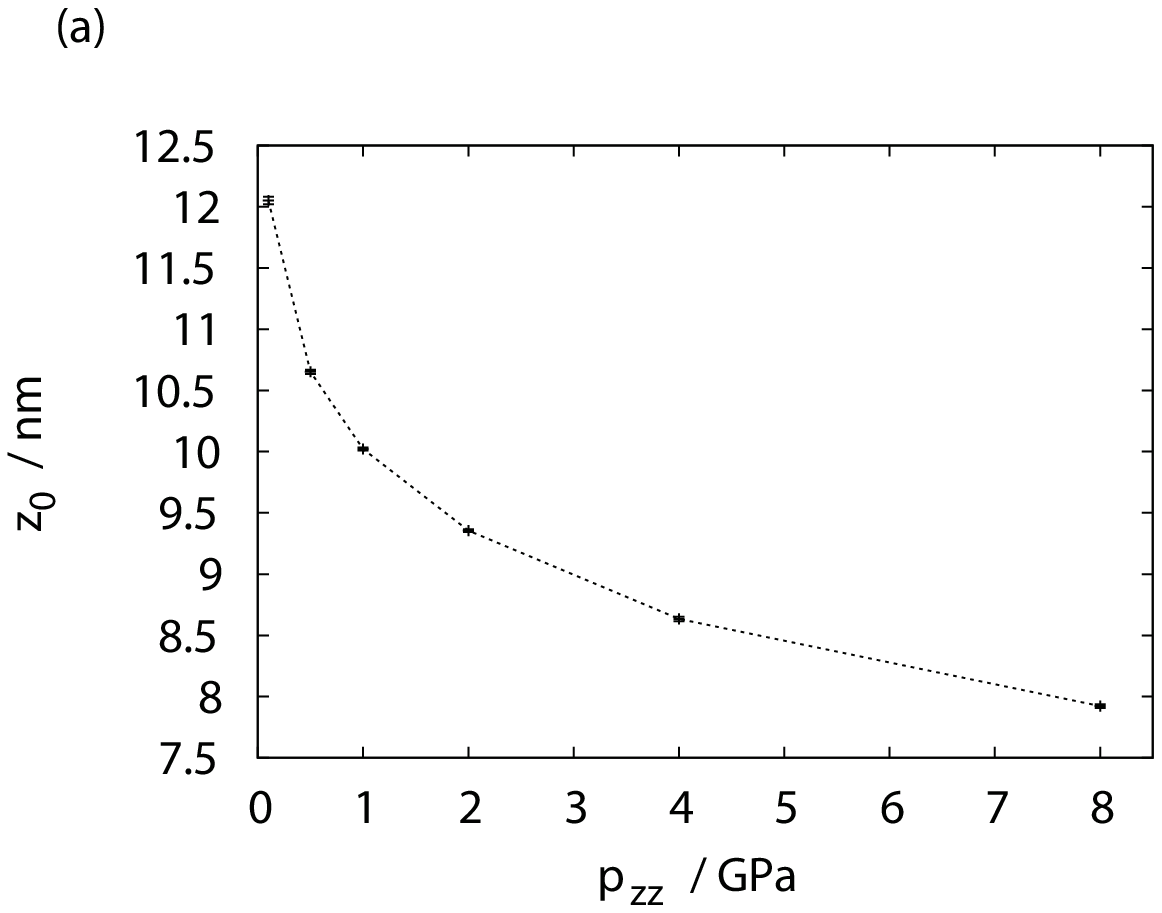}
 \includegraphics[width=80mm]{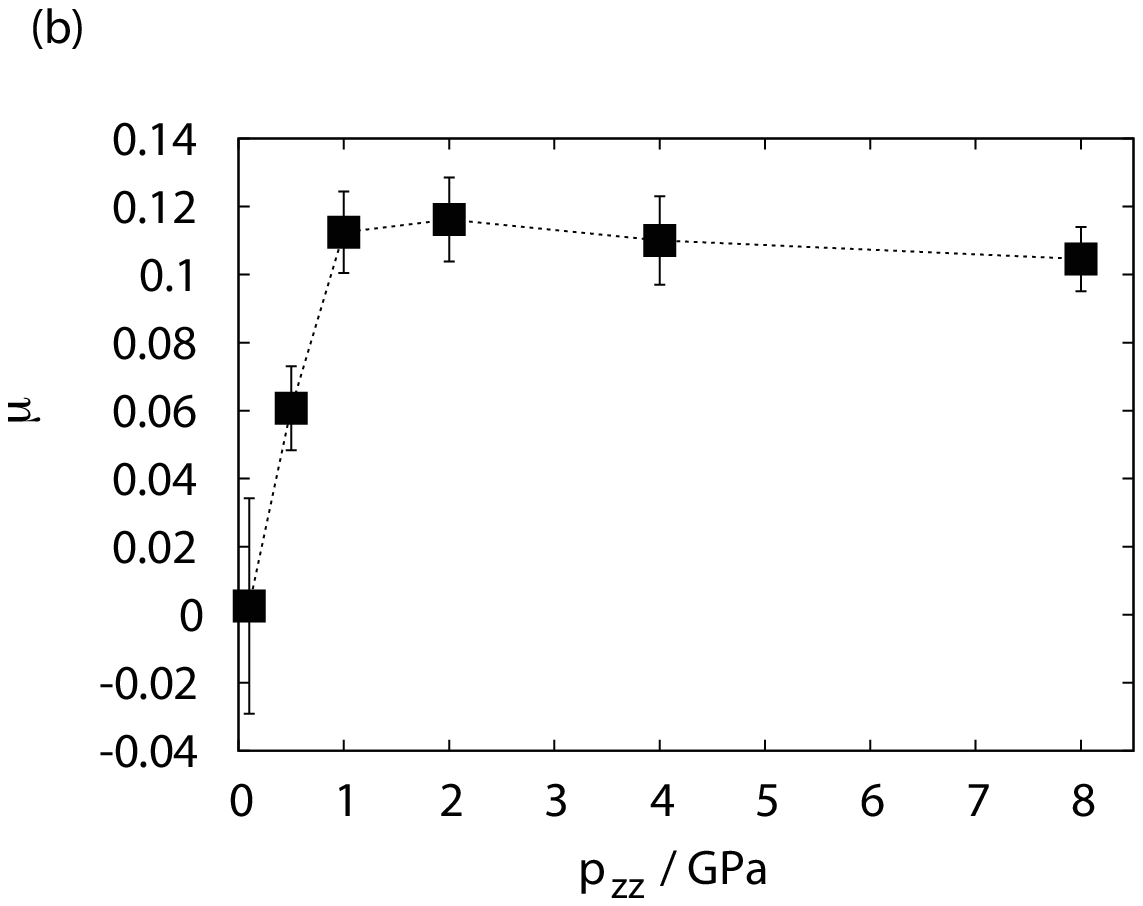}
 \caption{
 {
 External pressure dependencies of
 (a) the film thickness and
 (b) the traction
 coefficient calculated from MD simulations.
 }}
  \label{fig:p-mu2}
\end{center}
\end{figure}

\begin{figure}[h]
\begin{center}
 \includegraphics[width=80mm]{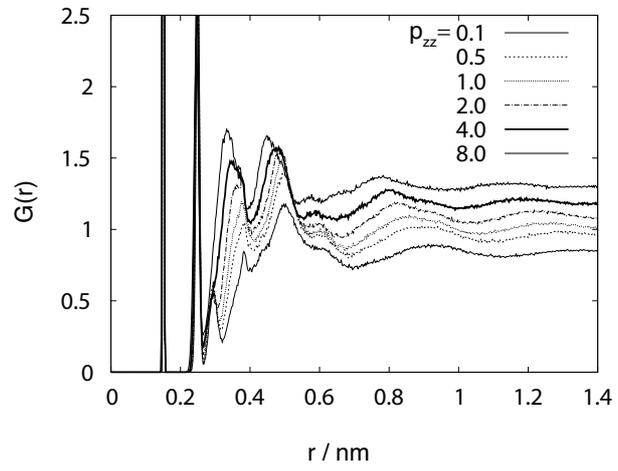}
 \caption{
 {
Radial distribution functions (RDFs) of carbon
atoms $G(r)$ at
various external pressures.
All lines are normalized by 
the average concentration at $p_{zz} = 1.0$ GPa.
 }}
  \label{fig:mksort10}
\end{center}
\end{figure}

\begin{figure}[h]
\begin{center}
 \includegraphics[width=80mm]{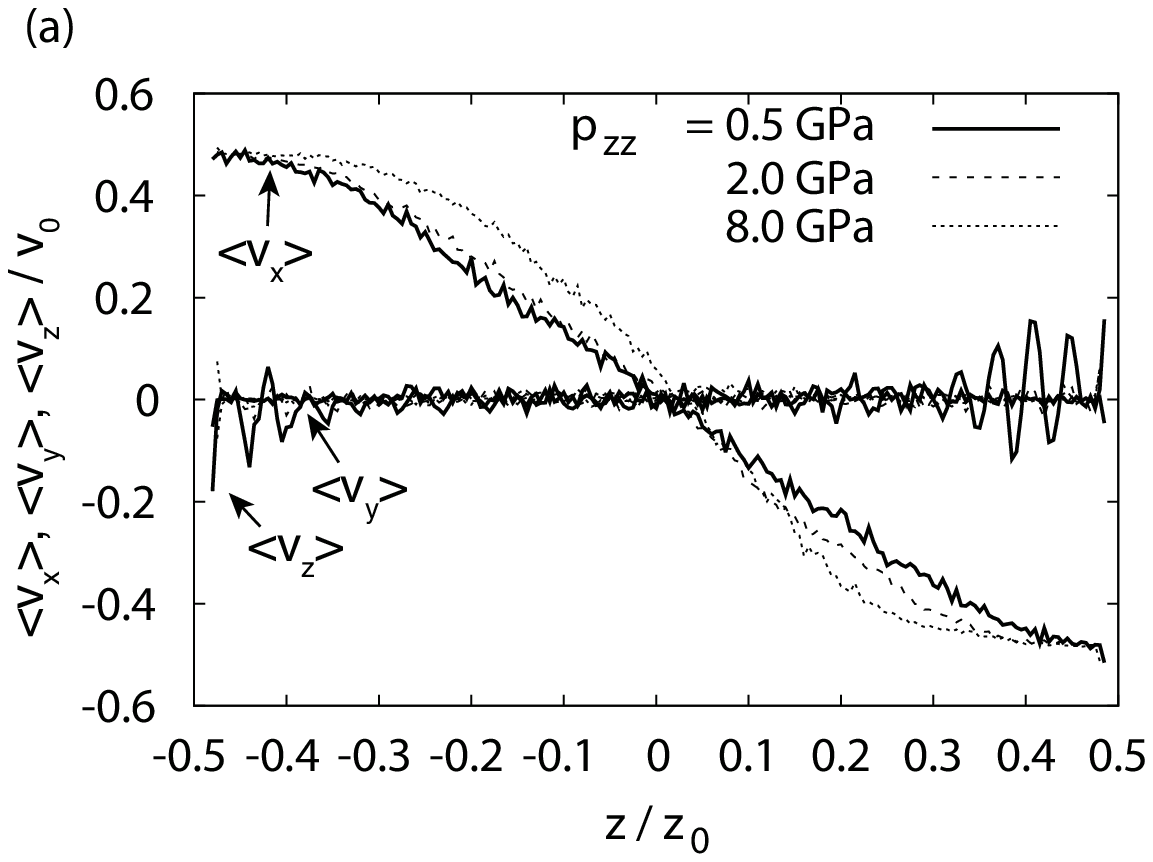}
\end{center}
\vspace{10mm}

\begin{center}
 \includegraphics[width=80mm]{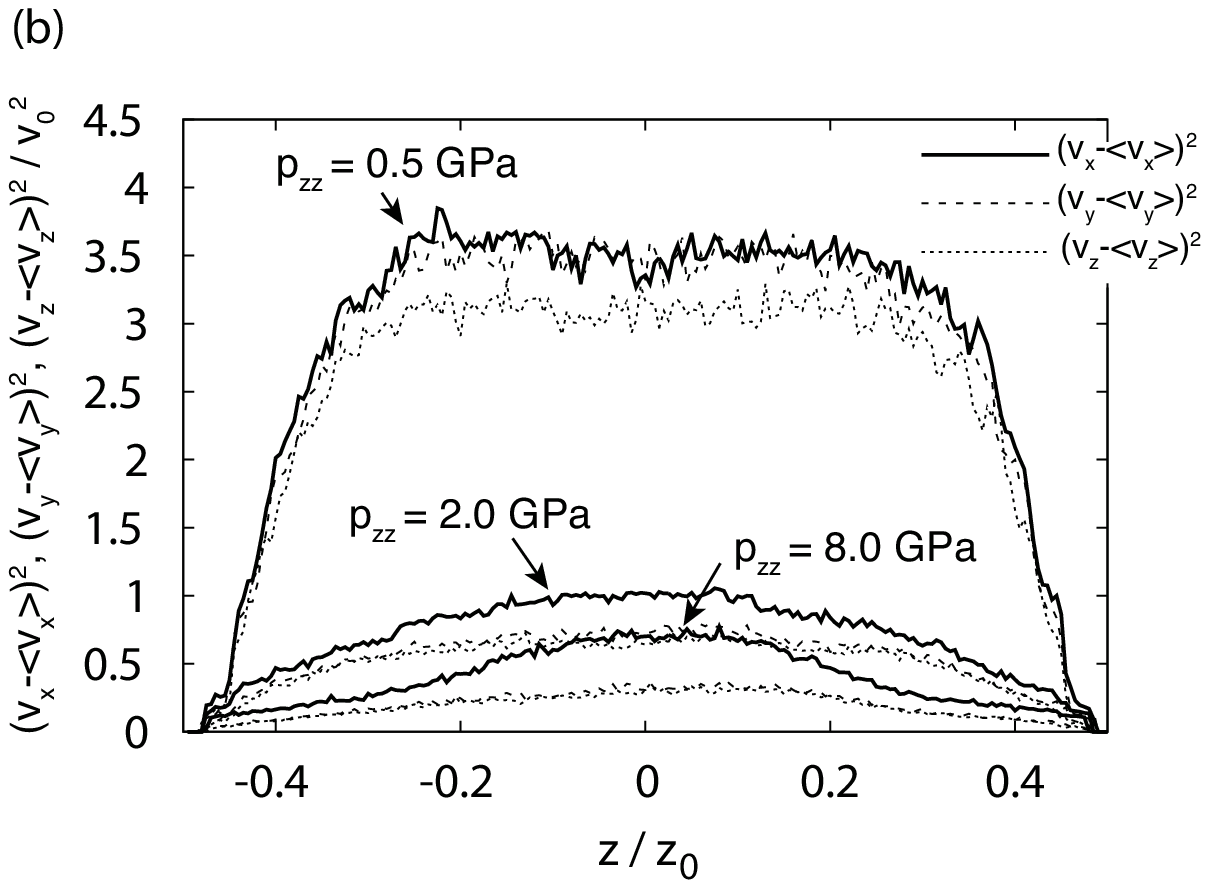}

 \caption{
 {
 Profiles of (a) the velocities
 $v_{x}, v_{y}$, and $v_{z}$ 
 and (b) the velocity fluctuations
 $(v_{x} - <v_{x}>)^2, (v_{y} - <v_{y}>)^2$, and $(v_{z} - <v_{z}>)^2$
at external pressures $p_{zz} = 0.5$, 2.0, and 8.0~GPa.
}}
 \label{fig:mksort6m2}
\end{center}
\end{figure}

\begin{figure}[h]
\begin{center}
 \includegraphics[width=80mm]{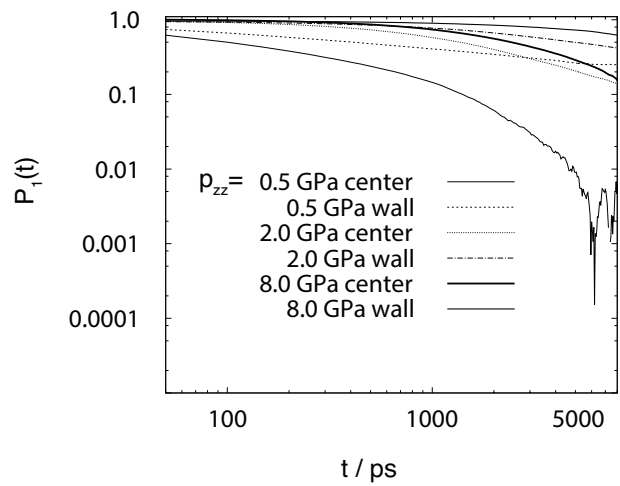}
 \caption{
 {
Variations in the segment correlation
functions $P1(t)$ 
 at external pressures
 $p_{zz} = 0.5$, 2.0, and~8.0 GPa.
}}
 \label{fig:mksort12a}
\end{center}
\end{figure}

\end{document}